\documentclass[letterpaper,11pt,onecolumn]{IEEEtran}
\usepackage{epsfig}
\usepackage[usenames]{color}
\usepackage[margin=1.02in]{geometry}
\usepackage[lined,algonl,boxed]{algorithm2e}
\usepackage{ifthen}
\usepackage{amsmath, amsthm, amssymb} 
\newtheorem{theorem}{Theorem}
\newtheorem{lemma}{Lemma}
\newtheorem{corollary}{Corollary}

\newtheorem{definition}{Definition}
\newcommand{\pr}[1]{\mbox{$\mathcal{P}\left[#1\right]$}}
\newcommand{\ev}[1]{\mbox{$\mathbb{E}\left[#1\right]$}}

\newcommand{\ora}[3]{\mbox{$\mathcal{O}(#1,#2,#3)$}}

\newcommand{\dbp}{\mbox{$\mathcal{T}$}}
\newcommand{\msp}{\mbox{$\mathcal{K}$}}
\newcommand{\mc}[1]{\mbox{$\mathcal{#1}$}}
\newcommand{\ball}[2]{\mbox{$\mathcal{B}_{#1}(#2)$}}
\newcommand{\balld}[2]{\mbox{$\beta_{#1}(#2)$}}
\newcommand{\mlf}[1]{
	\[
	\begin{array}{ll}
		#1
	\end{array}
	\]}
\newcommand{\brak}[1]{\left\{#1\right\}}
\begin{document}
\title{\vspace{20pt}Approximate Nearest Neighbor Search through Comparisons\\\vspace{50pt}}
\author{
\IEEEauthorblockN{Dominique Tschopp and Suhas Diggavi\\}
\IEEEauthorblockA{School of Computer and Communication Sciences\\
Ecole Polytechnique F\'ed\'erale de Lausanne (EPFL)\\
1015 Lausanne, Switzerland\\
Email: (first\_name.last\_name)@epfl.ch\\
}
}
\maketitle
\begin{abstract}
  This paper addresses the problem of finding the nearest neighbor (or
  one of the R-nearest neighbors) of a query object $q$ in a database
  of $n$ objects. In contrast with most existing approaches, we can
  only access the ``hidden'' space in which the objects live through a
  similarity oracle.  The oracle, given two reference objects and a
  query object, returns the reference object closest to the query
  object. The oracle attempts to model the behavior of human users, capable of making statements about similarity, but not of assigning meaningful numerical values to distances between objects. Using such an oracle, the best we can hope for is to obtain, for every object $u$ in the database, a sorted list of the other objects according to their distance to $u$. We call the position of object $v$ in this list the \emph{rank} of $v$ with respect to $u$. The difficulty of searching using such an oracle depends on the non-homogeneities of the underlying space. We use two different characterizations of the underlying space to capture this property. The first one, \emph{rank distortion}, relates pairwise ranks to the average difference in ranks w.r.t. other objects (a more precise definition is given in Section \ref{sec:def}). The second one, the combinatorial framework (a notion from \cite{318,320}), defines approximate triangle inequalities on ranks (a more precise definition is given in Section \ref{sec:def}). Roughly speaking, it defines a multiplicative factor $D$ by which the triangle inequality on ranks can be violated. Utilizing the insights from these ideas, we develop a
  hierarchical search algorithm that builds a data structure, which
  allows us to retrieve the nearest neighbor with high probability in
  $O(D^3\log^2 n\log\log n^{D^3})$ questions. The learning
  requires asking $O(n D^3\log^2 n\log\log n^{D^3})$ questions in
  total and we need to store $O(n\log^2n/\log(2D))$ bits in
  total. We also provide an approximate nearest neighbor search algorithm. Finally, we show a lower bound of $\Omega(D\log \frac{n}{D^2}+D^2)$ average number of questions
  in the search phase for randomized algorithms when the answers to all possible questions in
  the learning phase are given. We also introduce \emph{rank-sensitive} hash functions which
  gives same hash value for ``similar'' objects based on the
  rank-value of the objects obtained from the similarity oracle. As one application of RSH, we demonstrate that, we can retrieve one of the $(1+\epsilon)r$-nearest neighbor of a query point in $n^\theta$ evaluations of the hash function, where $\theta$ only depends on $\epsilon$ and the rank distortion.
\end{abstract}
\newpage

\section{Introduction}
Consider the situation where we want to search and navigate a database, but we do not know the underlying relationships between the objects. In particular, distances may be difficult to discern, or may not be
well-defined. Such situations are common with 
objects where human perception may be involved.  A collection of
pictures of faces, taken from different angles and distances is an illustration of such a dataset. Indeed, the distances between feature vectors might be far from
the similarity perceived by humans.  Notwithstanding, either with
human-assistance or approximate classification, we may be able to
determine the relative proximity of an object with respect to a small
number of other objects\footnote{We have implemented such a
  human-assisted system for a database of faces in a project called
  ``facebrowser'' \cite{fbrw:website}.}. Humans have the ability to compare objects and make statements about which are the most similar ones, though they can probably not
assign a meaningful numerical value to similarity. This led to the question of how to design search algorithms based on binary
similarity decisions of the type ``A looks more like B than C''.

More formally, we aim to design an algorithm that given a query object
(\emph{e.g., }a face), efficiently
returns an object that is similar to that object among the objects
in a database. To do so, we have access to a similarity oracle which, given two reference objects and a query object, can tell which
of the two reference objects is most similar to the query object. We measure the performance of all our algorithms in terms of the number of questions that we need to ask the oracle. We can pre-process the database during a learning phase, and use the resulting answers to facilitate the search process.

We do \emph{not} make the assumption that the ``hidden'' space in which the database objects live needs to be a metric space. Using this oracle one can retrieve for every object $u$ in the database, a sorted list of the other objects according to their distance to $u$. We call the position of object $v$ in this list the \emph{rank} of $v$ with respect to $u$, and denote it by $r_u(v)$. Clearly, this relationship can be asymmetric \emph{i.e., } $r_u(v)\neq r_v(u)$ in general. This setup raises several new questions and issues, as any space can be described by its ranks relationships. How much does the fact that the rank of some object $v$ w.r.t. some other object $u$ is $k$, and the rank of $w$ w.r.t. $u$ is $k'$ tell us about the rank of $w$ w.r.t. $v$? In this paper, we introduce the notion of \emph{rank distortion} (see Section \ref{sec:def} for a rigorous definition). The rank distortion captures how closely $r_v(w)$ is related to the average $\frac{1}{n}\sum_u |r_u(v)-r_u(w)|$. The framework introduced in \cite{320}, defines approximate triangle inequalities on the ranks, another way to capture these relationships. Those inequalities roughly tell us how ``transitive'' the similarity relationship is and give us a notion of \emph{combinatorial disorder}. If we have this information, we can use partial rank information to estimate, or infer the other ranks. In this paper, we will first investigate the case where we can use such a characterization of the hidden space as an input to our algorithms. We develop a randomized hierarchical scheme that improves the existing bounds for nearest neighbor search based on a similarity oracle (see Section \ref{sec:rew}).  We also prove, as far as we know, the first lower bound on the average number of questions to be asked for randomized nearest-neighbor search in this setup (see Section \ref{sec:skd}). Then, in Section \ref{sec:btd}, we ask what can be done if no characterization of the hidden space is known and therefore cannot be used as an input to the algorithms. In that case, we cannot estimate, or limit, ranks anymore if we have partial rank information. Nevertheless, we develop algorithms that can decompose the space such that dissimilar objects are likely to get separated, and similar objects have the tendency to stay together. This generalizes the notion of randomized $k$-$d$-trees (\cite{DasguptaRPT}) to our setup. Building on this intuition, we introduce the notion of \emph{rank-sensitive hashing} (RSH) in Section \ref{sec:rsh}. Similarly to locality-sensitive hashing, we can retrieve one of the $R$ nearest neighbors of a query point very efficiently. The hash function itself does not require any characterization of the subjacent space as an input. However, the smallest value of $R$ we can choose depends on the rank distortion. In general, both the criteria (combinatorial disorder and rank distortion) we use to characterize the hidden space seem to capture how ``homogeneous'' that space is. It appears that the less homogeneous it is, the more difficult it becomes to search. In particular, if the rank relationship is very asymmetric, and some objects are far from every other object, the information contained about those objects in the ranks matrix is very sparse and hard to capture. We apply this idea of RSH to NN search, but we believe that this might be useful in other scenarios as well.

\subsection{Relationship to published works}
\label{sec:rew}
The nearest neighbor (NN) problem, and many variations thereof, have
been extensively studied in the literature (see for instance \cite{clarksonsurvey} and \cite{indyksurvey} for surveys). In particular, very efficient algorithms have been developed for specific classes of metric spaces, such as metric spaces with a low intrinsic dimension or a bounded growth factor. In \cite{KrauthLee}, the authors introduce $\epsilon$-nets, a very simple data structure for nearest neighbor search (and many other applications). The complexity of those nets depends on the doubling dimension of the underlying space.   In \cite{KargerRuhl}, the authors present a random sampling algorithm to produce a data structure for search in growth restricted metrics. The restricted growth guarantees that a random sample will have some nice properties. In particular, by randomly selecting a small number of representatives at different scales for every object in a learning phase, one can zoom in on the nearest neighbor of a query point during the search phase. On the other hand, search when the underlying space is not necessarily a metric space appears to have very little prior work. In some sense, it is a generalization of the above problem, as any dataset can be represented by its rank relationships.  The problem of searching with a similarity oracle  was first studied in
\cite{320}\footnote{Our interest in this formulation arose from an
  applied viewpoint in the implementation of the facebrowser system
  \cite{fbrw:website}.}, where a random walk algorithm is presented. The main limitation of this algorithm is the fact that all rank relationships need to be known in advance, which amounts to asking the oracle $O(n^2\log n)$ questions, in a database of $n$ objects. The authors of \cite{318} and \cite{320}
work with a combinatorial framework for nearest neighbor search, which
defines approximates inequalities for ranks analogous to the triangle
inequality for distances. Their bounds depend crucially on the combinatorial disorder, represented by the
\emph{disorder constant} $D$ of the database (a notion to be defined more
formally in Section \ref{sec:def}, which captures to what extent the triangle inequality on ranks can be violated). In \cite{318}, a data structure similar in spirit to $\epsilon$-nets of \cite{KrauthLee} is introduced. It is shown that a learning
phase with complexity $O(D^7n\log^2n)$ questions and a space
complexity of $O(D^5n+Dn\log n)$ allows to retrieve the nearest
neighbor in $O(D^4\log n)$ questions, in a database of $n$ objects. The learning phase builds a
hierarchical structure based on coverings of exponentially decreasing
radii\footnote{the radius of a ball is defined as the cardinality of
  that ball.}. We will show (see Section \ref{sec:ctb}) that we can improve those bounds by a factor polynomial in $D$, if we are willing to accept a negligible (smaller than $\frac{1}{n}$) probability of failure. Our algorithm is based on random sampling, and hence can be seen as a form of metric skip list (as introduced in  \cite{KargerRuhl}), but applied to a combinatorial (non-metric) framework. However, the fact that we do not have access to distances forces us to use new techniques in order to minimize the number of questions we need to ask (or ranks we need to compute). In particular, we sample the database at different densities, and infer the ranks from the density of the sampling, which we believe is a new technique. We also need to relate samples to each other when building the data structure top down.  We also present what we believe is the first lower bound for our problem of searching through comparisons.

A natural question to ask is whether one can develop data structures for NN when a characterization of the underlying space is unknown. This has been addressed in the case when the underlying metric space has low "intrinsic" dimension and one has access to metric distances in \cite{KrauthLee, DasguptaRPT}. In \cite{DasguptaRPT}, it is shown that one can build a binary tree decomposition of a dataset of points in $\Re^d$, such that the diameter of the sets in the tree is reduced by a constant after a number of level that only depends on the intrinsic dimension of the data, and not $d$. The term intrinsic dimension either refers to the Assouad or the covariance dimension. Therefore, one can similarly ask such a natural question in our framework where we do not have access to metric distances (or they do not exist). We develop a binary tree (hierarchical) decomposition, when the characteristics of the underlying space (disorder constant) is unknown. This extends the result of \cite{DasguptaRPT} to our framework, where we only have access to the underlying space through comparisons.

The approximate nearest neighbor problem consists in finding an
element that is at distance at most $(1+\epsilon)d_{min}$ from the
query point $q$, where $d_{min}=\min_{i}d(i,q)$. In \cite{lshash},
Indyk and Motwani present two algorithms for this problem. 
In particular, \emph{locality sensitive hashing}, through which they obtain an algorithm with
polynomial learning and query time polynomial in $d$ and $\log
n$.  For binary vectors, it is remarkable that the performance of the algorithm does not depend on the dimension. A survey of results for LSH can be found in \cite{cacmLSH}. In \cite{322}, Panigrahy shows that instead of using a large
number of hash tables as it is the case in the approach above, only a
few can be used. These are then hashed to several randomly chosen
objects in the neighborhood of the query point, and it is shown shows
that at least one of them will fall into the same bucket as the
nearest neighbor. The authors of $\cite{313}$ prove a lower bound on the parameter
$\rho=\frac{\log{1/p}}{\log{p/P}}$ for $(r,cr,p,P)$-locality sensitive
hashing schemes. We present a new hashing scheme that is \emph{rank-sensitive} (RSH). How efficient the scheme is depends on another property of the hidden space, its \emph{rank-distortion}. The rank-distortion need not be an input to the algorithm, however, the performance will depend on it. We give sufficient conditions for RSH to work and demonstrate its application to NN search.

To the best of our knowledge, the notion of rank-sensitive hashing and
approximate (and randomized) nearest neighbor search using similarity
oracle is studied for the first time in this paper. Moreover, the hierarchical search scheme proposed is more efficient than earlier
schemes. The lower bound presented appears to be new and demonstrates
that our schemes are (almost) efficient.

\section{Definitions and Problem Statement}
\label{sec:def}
In this section, we define formally the notions that we use in the
rest of the paper. We consider a hidden space $\msp$ with distance function $d(.,.)$, and a database of objects $\dbp\subset\msp$, with $|\dbp|=n$. We do not have access to the distances between
the objects in $\msp$ directly. We can only access this space
through a \emph{similarity oracle} which for any point $q\in\msp$, and
objects $u,v\in\dbp$ returns:
\begin{equation}	
\ora{q}{u}{v}=
 \left\{
\begin{array}{ll}
 u & \mbox{if }d(u,q)\leq d(v,q)\\
 v & \mbox{else}
\end{array}
\right.
\end{equation}
For the sake of simplicity, we consider that all distances in $\msp$
are different. Note that the objects do not need to be in an underlying metric space for this similarity oracle. We now define the notion of \emph{rank}.
\begin{definition}
  The rank of $u$ in a set $\mc{S}$ with respect to $v$,
  $r_v(u,\mc{S})$ is equal to $c$, if $u$ is the $c^{th}$ nearest
  object to $v$ in $\mc{S}$.
\end{definition}
To simplify the notation, we only indicate the set if it is unclear
from the context \emph{i.e.,} we write $r_v(u)$ instead of
$r_v(u,\mc{S})$ unless there is an ambiguity. Note that rank need not be a symmetric relationship between objects \emph{i.e.,} $r_u(v)\neq r_v(u)$ in general. Further, note that we can rank $m$ objects w.r.t. an object $o$ by asking the oracle $O(m\log m)$ questions. To do so, create the ranking w.r.t. $o$ by adding one object at a time. Observe that in order to add the $(i+1)^{th}$ object to the list, we need to ask $\log(i)$ questions. More precisely, we need to ask whether the $(i+1)^{th}$ object is closer to $o$ than the object currently at position $i/2$. Then, we can recurse on the set new set of objects (\emph{e.g.}, if the object to insert is closer than the $i/2^{th}$ object, select the $i/4^{th}$ object as the new ``pivot''). Summing over $i$, the total number of questions to be asked to sort $m$ objects is $O(m\log(m))$.  

Our first characterization of the space of objects is through a form of approximate triangle
inequalities first introduced in \cite{318} and \cite{320}\footnote{We have another characterization called rank distortion in Definition \ref{def:rdi}}. Instead of
defining a relationship between distances, these triangle inequalities
define a relationship between ranks. These relationships depend on a
property of the space called the \emph{disorder constant}
$D$. In \cite{318} and \cite{320}, four such inequalities are defined,
all implying the others with $D'=D^2$.
\begin{definition}
\label{apxti}
The rank disorder of a set of objects $S$ is the smallest $D$ such that $\forall
x,y,z\in S$, we have the following approximate triangle
inequalities:
\begin{enumerate}
	\item $r_x(y,S)\leq D(r_z(x,S)+r_z(y,S))$
	\item $r_x(y,S)\leq D(r_x(z,S)+r_y(z,S))$
	\item $r_x(y,S)\leq D(r_x(z,S)+r_z(y,S))$
	\item $r_x(y,S)\leq D(r_z(x,S)+r_y(z,S))$
\end{enumerate}
In particular, $r_x(x,S)=0$ and $r_x(y,S)\leq Dr_y(x,S)$.
\end{definition}
We define a rank-ball of radius $r$ around some point $x$ as
$\ball{x}{r}=\left\{i\in S|r_x(i)\leq r\right\}$. A ball in distance is defined as $\balld{u}{r}=\left\{i\in S|d(u,i)\leq r\right\}$
\par
We further define the rank matrix $\mathcal{R}$ where $r_{ij}=r_i(j)$, and the matrix $\mc{W}=\mc{R}+\mc{R}'$ (note that the matrix $\mc{W}$ is symmetric). For a subset $S\in\msp$, we define its diameter $\Delta_S=\max_{i,j\in S}w_{ij}$. Let $\rho_i$ denote the $i^{th}$ column of $\mc{R}$ \emph{i.e.,} we associate to every object $o\in\dbp$ a vector $\rho_o\in \left\{0,...,n-1\right\}^n$, such that the $j^{th}$ coordinate of $o$ is given by $r_j(o)$. 
\par
We now define the \emph{rank-distortion} of a set $S$ as follows:
\begin{definition}
\label{def:rdi}
We say of a set of objects $S$ that its rank distortion function is $f:\mathbb{N}_+\rightarrow\mathbb{N}_+$, if $f$ is monotonically increasing and if there exists $\gamma>0$ (the \emph{rank-distortion}) such that $\forall u,v\in S$:
$$
f(r_u(v))\leq ||\rho_v-\rho_u||_1\leq \gamma f(r_u(v))
$$
\end{definition}
\begin{lemma} 
If the function $f$ is linear \emph{i.e.,} $f=cr_u(v)$, then the four approximate triangle inequalities are implied with $D\leq\gamma$. 
\end{lemma}
For example, for the first inequality, we have $r_x(y,\msp)\leq ||\rho_x-\rho_y||_1/c \leq (||\rho_x-\rho_z||_1+||\rho_z-\rho_y||_1)/c \leq \gamma(r_z(x,\msp)+r_z(y,\msp))$. The proof for the other inequalities is similar.

We can define the nearest neighbor problem as follows:
\begin{definition}[$R$-nearest neighbor problem]
  Given a set of objects $\dbp$ and a query point $q$, return one of
  the $R$ objects in $\dbp$ closest to $q$. In particular, if $R=1$,
  return the closest object to $q$ in $\dbp$.
\end{definition}
We say that a hashing scheme is $(r,R,p,P)$-sensitive if
\begin{definition}
  We call a hashing scheme $h$, "$(r,R,p,P)$-rank-sensitive" if
  $\forall q\in \msp, u\in\dbp$,
\[
\pr{h(q)=h(u)|r_q(u,\dbp)<r}>p\mbox{ and }\pr{h(q)=h(u)|r_q(u,\dbp)>R}<P
\]
\end{definition}
Note that we should have $P<p$. 

Finally, we say that a result holds \textbf{with high probability (w.h.p.)} if it hold with probability higher than $1-\frac{1}{n}$.

\section{Contributions}
\label{sec:ctb}
One of the difficulty of searching a hidden space arises from the fact that we cannot know how transitive the rank relationship is \emph{i.e.,} we cannot know whether the fact that A is similar to B, and B is similar to C implies that A is similar to C. This is problematic in the sense that even if the oracle tells us that A is closer to our query point than B, it does not necessarily imply that points close to A are better candidates than points close to B. In metric spaces, such a guarantee is provided by the triangle inequality. A way to characterize the hidden space is to limit the extent to which the triangle inequality on ranks can be violated. The combinatorial framework, introduced in \cite{318,320}, (see definition of approximate triangle inequalities) does exactly that. In this paper, we improve on their results in two ways. We provide more efficient algorithms using randomization and also provide a new lower bound for such randomized algorithms. More precisely, we show that if we only require success with high probability for nearest neighbor search, we can exploit the fact that a sample of randomly chosen points will have nice properties. In particular, it will be very likely that every object in the database will have an object sampled that is similar to itself. By sampling more and more densely at every level of a hierarchy, we will ultimately sample all objects. The key observation is that in order to find the sample closest to a particular object, we will only need to look at objects for which the closest sample at the level above in the hierarchy was also close to that object. We introduce a conceptually simple randomized hierarchical scheme that
allows us to reduce the learning compared to the existing
algorithm (see \cite{318,320}) by a factor $D^4$, memory consumption by a factor
$D^5/\log^2 n$, and a factor $D/\log n\log\log n^{D^3}$ for search
(see Section \ref{sec:rew}). This algorithm's performance is best
when the disorder constant is small. 
\begin{theorem}
\label{thm:ubo}
There exists a data structure, which for a given query point $q$, can
retrieve its nearest neighbor with high probability in $O(D^3\log^2
n\log\log n^{D^3})$ questions. The learning requires asking $O(n
D^3\log^2 n\log\log n^{D^3})$ questions in total. We need to store
$O(n\log^2n/\log(2D))$ bits in total.
\end{theorem}
We then prove a lower bound on the average search time to retrieve the nearest neighbor of a query point for randomized algorithms. Our result confirms the intuition we have developed so far. Indeed, the higher the disorder constant, the more difficult it becomes to search. One way to interpret this result is that the higher the disorder constant D, the less information the answer to a question to the Oracle provides us. 
\begin{theorem}
\label{thm:lbo}
There exists a space, a configuration of a database of $n$ objects in that space and a distribution over placements of the query point $q$ such that no randomized search algorithm,
even if $O(n^3)$ questions can be asked in the learning phase,
can find $q$'s nearest neighbor in the database for sure (with a probability of error of 0) by asking less than an expected $\Omega(D\log(\frac{n}{D^2})+D^2)$ questions.
\end{theorem}
Consequently, our schemes are asymptotically (for $n$) within a factor $\tilde{O}(D)$ of the optimal scheme (\emph{i.e.,} within $D polylog(n)$ questions of the optimal search algorithm). The proofs of those two theorems are provided in Section \ref{sec:skd}.
\par 
Clearly, one of the limitations of the schemes above is that we need to know the disorder constant. It might be possible to estimate the value of the disorder constant based on a sample of objects in the database. Limitations of this approach are the fact that we might considerably degrade the performance of the algorithms if the estimator is inaccurate, and that we might run into trouble if the query point does not come from the same distribution as the database points $\dbp$. We therefore extend, in Theorem \ref{thm:btp}, the idea of $k$-$d$-trees to our setup. We provide an algorithm to build a binary tree that adapts to the disorder of the hidden space  (see \cite{DasguptaRPT} for an analogous result for $\Re^d$).   
In Section \ref{sec:rsh}, we present a new \emph{rank-sensitive} hash function with many potential applications. The idea of rank-sensitive hashing is that by computing many times a hash function drawn at random, similar objects will be assigned the same hash value more frequently that dissimilar objects. The performance of the rank-sensitive hashing scheme depends on the rank-distortion of the hidden space. Instead of capturing how ``transitive'' the rank relationship is, the rank disorder captures how the rank $r_{u}(v)$ relates to the average rank \emph{i.e.,} $\ev{|r_{j}(v)-r_j(u)|}$. In other words, if we picked an object $x$ at random, and sorted all other objects w.r.t. this object, how would $|r_{x}(v)-r_x(u)|$ relate to $r_u(v)$? If $r_u(v)$ can be approximated by a function $f$ of $\ev{|r_{j}(v)-r_j(u)|}$), then we can exploit this fact to separate points close to $q$ and points far from $q$. 
\begin{theorem}
\label{thm:rhf}
Given a set of objects $S$ with rank-distortion function $f$, and rank distortion $\gamma$, there exists a function $h$ which is $(r,(1+\epsilon)r,1-\frac{f(r)}{n^2},1-\frac{f((1+\epsilon)r)}{n^2\gamma })$-rank-sensitive.
\end{theorem}
A special case is when the function $f$ is constant. Then, the behavior of the function is similar to the one observed with locality-sensitive  hashing for binary vectors. One of the consequences is that we can retrieve one of the $R=(1+\epsilon) r$ nearest neighbors of a query point $q$ in $n^{O(\frac{\gamma}{\epsilon})}$ questions. By using the output of the hash function in a different way, we can compute an overall ranking of the objects. We can then retrieve "popular" objects \emph{i.e.,} those which are close to many other objects. This idea is discussed in Section \ref{sec:btree}.

\section{Searching with Known Disorder Constant}
\label{sec:skd}
In this section, we make the assumption that the disorder constant $D$, of $\dbp\cup\left\{q\right\}$ is known, and that we can consequently use it as an input to our algorithms. Knowing $D$ is an advantage, as it allows one to rapidly exclude some candidate objects during the search phase. In other words, we can take advantage of the fact that if we found an object close to the query point $q$, then objects which are far from that object cannot be the nearest neighbor of $q$. We first present an algorithm for nearest-neighbor search. The algorithms builds a hierarchical decomposition of the test set $\dbp$. The construction succeeds with high probability \emph{i.e.,} for a fixed query point $q$, the data structure is such that it will return $q$'s nearest neighbor w.h.p. Then, we present a lower bound on the search complexity.

\subsection{Hierarchical Data Structure For Nearest-Neighbor Search}
\label{sec:algh}
The learning phase is described in Algorithm \ref{alg:prp}. The algorithm builds a hierarchical decomposition level by level, top-down. At each level, we sample objects from the database. The set of samples at level $i$ is denoted by $S_i$, and we have $|S_i|=m_i=a(2D)^i\log n$, where $a$ is a constant independent of $n$ and $D$. At each level $i$, every object in $\dbp$ is put in the ``bin'' of the sample in $S_i$ closest to it. To find this sample at level $i$, for every object $o$ we rank the samples in $S_i$ w.r.t. o (by using the oracle to make pairwise comparisons). However, we will show that given that we know $D$, we only need to rank those samples that fell in the bin of one of the at most $4aD\log n$ nearest samples to $o$ at level $i-1$. This is a consequence of the fact that we carefully chose the density of the samples at each level. Further, the fact that we build the hierarchy top-down, allows us to use the answers to the questions asked at level $i$, to reduce the number of questions we need to ask at level $i+1$. This way, the number of questions per object does not increase as we go down in the hierarchy, even though the number of samples increases. The search process is described in Algorithm \ref{alg:sea}. The key idea is that the sample closest to the query point on the lowest level will be its nearest neighbor. Hence, by repeating the same process as for inserting objects in the database, we can retrieve the nearest neighbor w.h.p.
\incmargin{1em}
\restylealgo{boxed}\linesnumbered
\begin{algorithm}
\label{alg:prp}
\SetKwData{Left}{left}
\SetKwData{This}{this}
\SetKwData{Up}{up}
\SetKwFunction{Union}{Union}
\SetKwFunction{FindCompress}{FindCompress}
\SetKwInOut{Input}{input}
\SetKwInOut{Output}{output}
\caption{Learning Algorithm}
\Input{A database with $n$ objects $z_1,...,z_n$ and disorder $D$} \Output{For each
  object $u$, a vector $\phi_u$ of length $\log n/\log(2D)$. The list
  of all samples $\cup_iS_i$} \BlankLine \For{$i\leftarrow 1$ \KwTo $L=\frac{\log
    n}{\log 2D}$}{ \emph{Let $S_i$ be a set of $a(2D)^i\log n$ objects
    chosen u.a.r. in the database \dbp}\;    
    \For{$j\leftarrow 1$ \KwTo $n$}
    {\nllabel{forins}
     \eIf{$i=1$}{$c_j(1)\leftarrow
    S_1$} 
{
$c_j(i)\leftarrow \left\{v\in
    S_{i}|\mbox{position of $\phi_v(i-1)$ in $c'_j(i-1)$ smaller than $4aD\log n$}\right\}$\; 
\tcc{$c_j(i)$ is the set of samples in $S_i$, for which the closest sample in $S_{i-1}$ was one of the (at most) $4aD\log(n)$ closest sample to $z_j$ in $S_{i-1}$}
  }  
      \uIf{$|c_j(i)|=0$}
      {Report Failure}
      \Else{
      $c'_j(i)\leftarrow$ sort $c_j(i)$ according to $r_{z_j}(v,S_i)$, $\forall v\in c_j(i)$\;
    	$\phi_j(i)\leftarrow$ first object in $c'_j(i)$\;  \tcc{$\phi_j(i)$ is the sample in $S_i$ closest to $z_j$}
	} }
			}
\end{algorithm}
\decmargin{1em}

\incmargin{1em}
\restylealgo{boxed}\linesnumbered
\begin{algorithm}
\label{alg:sea}
\SetKwData{Left}{left}
\SetKwData{This}{this}
\SetKwData{Up}{up}
\SetKwFunction{Union}{Union}
\SetKwFunction{FindCompress}{FindCompress}
\SetKwInOut{Input}{input}
\SetKwInOut{Output}{output}
\caption{Search Algorithm}
\Input{A database with $n$ objects and disorder $D$, the list of
  samples, the vectors $\phi$, a query point $q$} \Output{The nearest
  neighbor of $q$ in the database} 
   \BlankLine 
  $c'_q(1)=S_1$\;
 \For{$i\leftarrow 2$
  \KwTo $L=\frac{\log n}{\log 2D}$}{ $c_q(i)\leftarrow \left\{v\in
    S_{i}|\mbox{position of $\phi_v(i-1)$ in $c'_q(i-1)$ smaller than $4aD\log n$}\right\}$\;
  $c'_q(i)\leftarrow$ sort $c_q(i)$ according to $r_{q}(v,S_i)$, $\forall v\in c_q(i)$\; } \Return first object in $c'_q(\frac{\log
  n}{\log 2D})$
\end{algorithm}
\decmargin{1em}
We will now show that Algorithm \ref{alg:prp} succeeds with probability higher than $1-\frac{1}{n}$ (w.h.p.) and that it requires asking less than $O(D^3\log^2
n\log\log n^{D^3})$ questions w.h.p.
\begin{theorem}
\label{thm:alg1}
Algorithm \ref{alg:prp} succeeds with probability higher than $1-\frac{1}{n}$ (w.h.p.) and it requires asking less than $O(nD^3\log^2n\log\log n^{D^3})$ questions w.h.p.
\end{theorem}
\begin{proof}
See Appendix  \ref{apx:alg1}
\end{proof}
The proof of Theorem \ref{thm:ubo} is then immediate and is given in Appendix \ref{apx:ubo}. Note that this scheme can be easily modified for $R$-nearest neighbor search. At the $i^{th}$ level of the hierarchy, the closest sample to $q$ will, w.h.p., be one of its $\frac{n}{(2D)^i}$ nearest neighbors. If we are only interested in the level of precision, we can consequently stop the construction of the hierarchy at the desired level.

\subsection{Lower Bound}
\label{sec:lbo}
In this section, we show that there exists configurations of $n$
objects in a graph metric for which no search algorithm can be
guaranteed to find the nearest neighbor of a query point in less than
expected $\Omega(D\log\frac{n}{D^2}+D^2)$ questions. We make the assumptions that all possible
questions related to the $n$ database objects can be asked during the
learning phase, and even that the structure of the database is
known. Then, we attach a query point to the database constellation in
a random way. Consider the graph shown in Fig. \ref{fig:lbt}. It is a
star with $\alpha$ branches, each composed of $n/\alpha^2$
\emph{supernodes}. All edges in this part of the graph have weight
$1$. Inside each supernode, there are $\alpha$ database objects. A \emph{root} node that
connects the supernode to the other supernodes, and $\alpha$ objects,
each connected to the root with a different edge. The weights
of these edges range from $1/4\alpha$ to $\alpha/4\alpha$. Finally, the query point
will be connected to one object on every branch of the star. Hence, the
query point has $\alpha$ \emph{direct neighbors} (one on each branch of the star). The edges connecting the query
point to the graph have weights ranging from $1$ to
$1+\epsilon$, where $\epsilon\ll 1/4\alpha$. Note that we cannot know which are the direct neighbors of
the query point, nor what the weights of the corresponding edges
are. Thus, given the $n$ database objects and the answers to all
possible questions we can ask about the database, we need to find the
nearest neighbor of the query point. First, we show that this structure has disorder $\Theta(\alpha)$ (proof in Appendix  \ref{apx:dlb}).
\begin{lemma}
\label{lem:dlb}
  The graph shown in Fig. \ref{fig:lbt} with the shortest path
  distance has disorder constant $D=\Theta(\alpha)$.
\end{lemma}
In the proof of Theorem \ref{thm:lbo} (see Appendix \ref{apx:lbo} for a full proof), we show that we can lower bound the expected running time of any randomized algorithm on the example of Figure \ref{fig:lbt}. The idea of the proof is that we must identify and compare all direct neighbors to the query point, and then find the nearest neighbor among the direct neighbors. We show that we cannot identify all direct neighbors in fewer than an expected $\Omega(D\log \frac{n}{D^2}+D^2)$ questions.

\section{Searching with Unknown Characterization}
\label{sec:btd}
When the disorder constant is unknown, we cannot be sure to retrieve the nearest neighbor of a query point $q$, unless we ask $O(n)$ questions\footnote{unless we go sequentially through all objects and compare them to the current nearest neighbor, there could always be an object closer to the query point.}. In Section \ref{sec:skd}, we heavily relied on the fact that we could find objects close to the query point. Knowing the disorder constant then allowed us to exclude other objects as near neighbors. If we do not know $D$, we can still hope that by building a hierarchical decomposition, dissimilar objects will be separated rapidly as we walk down from the root to a leaf. Hence, we would also expect objects similar to $q$ to be close to it in the tree. However, we cannot bound this distance, as we cannot use $D$ as an input to the algorithm. First, however, we will give an example of a simple and intuitive algorithm that shows how much we can gain by knowing the disorder constant.   

\subsection{Consequences of knowing the disorder constant}
\label{sec:ckd}
If we know that an object
$u$ is the $j^{th}$ nearest neighbor of an object $x$ (\emph{i.e.,} we
have $r_x(u,\dbp)=j$), and we are looking for an object $y$, such that
$r_u(y,\dbp)<\zeta$, then we know that $y$ must lie in an annulus centered
at $x$ of a certain width around $u$ \emph{i.e.,} we know that $a\leq
r_x(y)\leq b$, where $a$ and $b$ are functions of $j$ and $\zeta$. In
particular, we have (proof in Appendix \ref{apx:als}):
\begin{lemma}
\label{lem:als}
Consider three objects $x$, $y$, and $u$. Let $r_x(u)=j$ and $r_y(u)<
\gamma$ (1), or $r_u(y)<\zeta$ (2). Then, $y$ must lie in an annulus such that
$\frac{j}{D}-\zeta\leq r_x(y)\leq D(j+\zeta)$.
\end{lemma}   
By sampling $m$ objects u.a.r., and computing all ranks w.r.t. to these objects, we can thus narrow down the search space to an annulus of width depending on $D$ and on the rank of the closest sample (proof in Appendix \ref{apx:ann}). 
\begin{theorem}
\label{thm:ann}
  Given a query object $q\in\msp$, we can retrieve one of its R nearest-neighbors in $\dbp$ by asking an expected $m+\log n+D+\frac{D^2n}{mR}+1$ questions, with constant probability. The learning phase requires asking an expected $O(nm\log n)$ questions to sort all objects w.r.t. the $m$ samples.
\end{theorem} 
In particular, by setting $m=\sqrt{\frac{n}{R}}$, we can retrieve one of the
$R$ nearest neighbors with constant probability in expected 
$O(D^2\sqrt{\frac{n}{R}})$ questions. The example is similar to what happens on a given level in the hierarchical scheme of Section \ref{sec:algh}. The fact that we know $D$, as is illustrated by the algorithm above, allows us to exclude some objects as being nearest neighbors. Indeed, if we have information about the rank of the sample w.r.t. the query point, or vice-versa, then we know the nearest neighbor must lie in an annulus of known width. On the other hand, if the disorder is unknown, we cannot exclude any object, whatever the density of the sampling. In the next section, we ask whether we can build a data structure that adapts to the characteristics of the space, without requiring it them as input. In other words, we ask whether we can decompose the space in such a way that dissimilar objects are likely to be separated, and similar objects remain close to each other, without knowing a characterization of the space.

\subsection{Binary Tree Decomposition}
\label{sec:btree}
A natural and simple way to build a data structure suited for search operations is to build a tree. By recursively applying Algorithm \ref{alg:rbc}, we can decompose the database into a binary tree. Clearly, this algorithm does not require any characterization of the space as an input. As illustrated in Section \ref{sec:ckd}, if we do not know $D$, we cannot limit the ranks if we only have partial rank information. However, we can expect this decomposition to adapt to the structure of the underlying space.
\incmargin{1em}
\restylealgo{boxed}\linesnumbered
\begin{algorithm}
\label{alg:rbc}
\SetKwData{Left}{left}
\SetKwData{This}{this}
\SetKwData{Up}{up}
\SetKwFunction{Union}{Union}
\SetKwFunction{FindCompress}{FindCompress}
\SetKwInOut{Input}{input}
\SetKwInOut{Output}{output}
\caption{Rank-Ball Cut}
\Input{A set S of objects $\in\dbp$} \Output{Two sets of objects $S_0$ and $S_1=S\backslash S_0$} \BlankLine 
pick two objects $x_1$ and $x_2$ u.a.r. in $S$\;
$S_0=\emptyset$, $S_1=\emptyset$\;
\ForAll{$u\in S$}
{
\lIf{$\ora{x_1}{x_2}{u}=u$}{$S_0=S_0\cup u$} \lElse{$S_1=S_1\cup u$}\;
}
 \end{algorithm}
\decmargin{1em}
Let the expected diameter after the decomposition of $S$ into $S_1$ and $S_0$ be defined as $\tilde{\Delta}_S=\frac{|S_0|}{|S|}\Delta_{S_0}+\frac{|S_1|}{|S|}\Delta_{S_1}\leq \Delta_S$ (by analogy to the notion in \cite{DasguptaRPT}). Observe that the diameter of a set (see definition in Section \ref{sec:def}) has the following property (proof in the Appendix \ref{apx:dia}).
\begin{lemma}
\label{lem:dia}
The diameter of a set $S$ with $|S|=n$ is always less than or equal to $2n$ \emph{i.e.,} $\Delta_S\leq 2n$, with equality when $d(u,v)=d(v,u)$ in the hidden metric space (symmetric distance function).
\end{lemma} 
We can compute the expected diameter after the decomposition of $S$ into $S_0$ and $S_1$. First, observe that it will always decrease, as the cardinality of the two new sets must be smaller or equal to the diameter of $S$. Let us denote by $x_1$ and $x_2$ the two randomly selected points in the set $S$. Let $r_{x_1}(x_2)=k$. By the approximate triangle inequality (1), for any pair of points $u$ and $v$ in $S_0$, we have 
$r_{u}(v)\leq D(r_{x_1}(u)+r_{x_1}(v))\leq 2Dk$.
Hence, the diameter $\Delta_{S_0}$ must be smaller or equal to $4Dk$. We can then easily compute the expected diameter to be $\tilde{\Delta}_S\leq \frac{4D}{n}k^2+2n-2k$.
Further, the optimal value for $k$ is $k=\frac{n}{4D}$. However, by choosing $x_2$ at random, we cannot ensure that $r_{x_1}(x_2)$ takes a specific value. Nevertheless, we know that the value of $k$ is uniformly distributed between $1$ and $n$. Assume that we want $\tilde{\Delta}_S\leq \epsilon 2n$, for some $\epsilon<1$. Then, we can prove the following theorem (proof in Appendix \ref{thm:btp}):
\begin{theorem}
\label{thm:btp}
Let $\epsilon<1$. Then,  $$\pr{\tilde{\Delta}_S\leq  \epsilon 2n}=\frac{1}{2D}\sqrt{1-8D(1-\epsilon)}$$
\end{theorem} 
Let a "good cut" be a cut such that the diameter is reduced by epsilon. The probability that we reduce the diameter y a factor $\epsilon$ degrades with increasing values of $D$. Hence, even though the disorder constant is not an input to the algorithm, the performance will depend on the disorder constant. For instance, if $D$ were constant, then we would reduce the diameter by a constant with constant probability. In general, we roughly need $\frac{\log(1/c)}{\log(\epsilon)}$ good cuts to divide the diameter by a constant $c$.  In any case, the depth of the binary tree is $O(\log n)$ w.h.p. (proof in the Appendix  \ref{apx:dbt}). An interesting fact is that the probability that a node $u$ falls in the good set \emph{i.e.,} the ball around $x_1$ is given by $\phi_u=\pr{u \in \ball{x_1}{r_{x_1}(x_2)}}=\frac{1}{n}\sum_j(1-\frac{r_j(u)}{n})$. Hence, "outliers" are likely to be put in the same bin as other outliers, while similar objects are likely to be put in the same bin. For instance, an object $y$ far away from all other objects, such that $\forall u\in\dbp$ we have $r_u(y)=n$, will hardly ever be put in the good set. Conversely, if there is a set of very popular nodes, which have a low rank w.r.t. all other objects, they will often end up in the good set. Consequently, this function can be used to estimate how ``central'', or popular an object is (analogous to the notion of 1-median in \cite{indykMedian}). $Y_i=\sum_{j} 1_{\left\{\mbox{node $i$ is in the good set}\right\}}$, where the sum goes over randomly selected hash functions and $1_{\left\{\right\}}$ is the indicator function, will provide such an estimate. It also implies that outliers are more likely to be separated from other objects. In particular, if we computed $k$ times the result of a randomly chosen hash function $h$, $Y_u$ would be roughly equal to $k\phi_u$. By sorting the $Y_i$'s, we obtain a ranking of the objects by popularity. In the next subsection, we will try to exploit this property to design a hashing scheme.

\subsection{Rank-Sensitive Hashing}
\label{sec:rsh}
We have developed the intuition that by randomly cutting out balls, it is more likely that similar objects will stay together, and dissimilar objects be separated. This should be sufficient, if we can amplify this property, to allow us to efficiently search for similar objects. Indeed, we will now show how we can use this technique to develop a rank sensitive hashing scheme. The rank distortion provides us a sufficient condition for the scheme to work. Our hash function $h$ selects two objects u.a.r in $\dbp$ (say $x_1$ and $x_2$), and assigns values $h(u)\in\left\{0,1\right\}$ to all objects $u$ as follows
$$
h(u)=\left\{\begin{array}{ll}0& \mbox{if }\ora{x_1}{x_2}{u}=u\\1  & \mbox{if }\ora{x_1}{x_2}{u}=x_2\end{array}\right.
$$
Note that computing $h$ requires asking a single question per object, and that the algorithm does not require any characterization of the space as input. The function $h$ is  $(r,(1+\epsilon)r,1-\frac{f(r)}{n^2},1-\frac{f((1+\epsilon)r)}{n^2\gamma })$-rank-sensitive. This is the result of Theorem \ref{thm:rhf}, proved in Appendix \ref{apx:rhf}. 
A special case is when the function $f$ is linear. Then, we obtain the following result  (proof in Appendix \ref{apx:rsh}).
\begin{corollary}
\label{cor:rsh}
We can retrieve one of the $(1+\epsilon)r$-nearest neighbors in $\dbp$ of a query point $q$, with constant probability, by asking $n^{O(\frac{\gamma}{\epsilon})}$ questions, where $\gamma$ is the rank distortion of $\dbp$, when the rank distortion function is linear.
\end{corollary}
Intuitively, one situation where $f$ is roughly constant is when the underlying space is close to a line in $\Re^d$. Further, our numerics have shown that even for higher dimensions, when the underlying space is homogeneous (\emph{e.g.,} points distributed u.a.r. in a unit box with wrap around distances), the function $f$ is very steep for small values of $r_u(v)$ and then almost linear. An example is given in Figure \ref{fig:rdu} in Appendix \ref{apx:nex}.

\section{Conclusions}
We addressed the problem of finding an object similar to a query
object among the objects in a large database. In contrast to most
existing formulations, we asked whether the database can be searched
efficiently if its distance information can only be accessed through a
similarity oracle, and the underlying objects need not be in a metric space. The oracle is motivated by a human user who can
make comparisons between objects but not assign meaningful numerical
values to similarities between objects. This raises new interesting questions on what are good properties of the rank relationships, what are good and efficient algorithms and what is the right characterization of such a space. We worked with two such characterizations in this paper. One that captures the transitivity of the rank relationship through disorder constant ($D$), and the other one, rank distortion ($\gamma$), which captures how rank $r_u(v)$ relates to $\mathbb{E}_i\left[r_i(v)-r_i(u)\right]$. We presented a new randomized algorithm that improves the performance of existing algorithms for the combinatorial framework, and proved a lower bound on the search complexity. We also propose a new characterization of the hidden space, \emph{rank-distortion}, and show that the performance of a novel rank-sensitive hashing scheme depends on that property. Rank-sensitive hashing enables (approximate) nearest
neighbor search in a manner similar to locality sensitive hashing. We believe that ideas of searching through comparisons form a bridge between many well known search techniques in metric spaces to perceptually important (non-metric spaces) situations.

\newpage

\bibliography{hms}

\appendix

\subsection{Proof of Theorem \ref{thm:alg1}}
\label{apx:alg1}
We first prove two technical lemmas that we will need to prove Theorem \ref{thm:alg1}. 
\begin{lemma}
\label{lem:tl1}
If we throw $m=ab\log n$ balls into $b$ bins, each chosen uniformly at
random, then the first bin will contain at least one ball with probability
$1-\frac{1}{n^a}$
\end{lemma}
\begin{proof}
  The probability that a bin contains no ball is \mlf{
    \pr{\mbox{a bin contains no ball}}&=(1-\frac{1}{b})^{ab\log n}\\
    &\leq e^{-a\log n}\\
    &=\frac{1}{n^a} } 
\end{proof}
\begin{lemma}
\label{lem:tl2}
We throw $m$ balls into $n$ bins, each chosen uniformly at
random. We number the bins from $1$ to $n$. Then, the probability that
the bins $1$ to $\frac{n}{c}$ contain more than $(1+\tau)m/c$ or less than $(1-\tau)m/c$ balls
is at most $2 e^{-\tau^2 m/3c}$.
\end{lemma}
\begin{proof}
  We throw the balls one after the other into the bins. Let $X_i=1$ if
  the $i^{th}$ ball falls in one of the $\frac{n}{c}$ first bins, and
  $0$ else. Let $X=\sum_i X_i$. Clearly, we have $\ev{X}=m/c$, as $\pr{X_i=1}=1/c$ and all $X_i$'s are independent. By the Chernoff Bound (see for instance \cite{Mitzbook}, page 67), we have  
  $\pr{|X-\ev{X}|>\tau m/c}<2 e^{-\tau^2 m/3c}$. 
\end{proof}
We are now ready to prove Theorem \ref{thm:alg1}
\begin{proof}
Let $m_i=a(2D)^i\log n$ denote the number of objects we sample at
level $i$, and let $S_i$ be the set of samples at level $i$ \emph{i.e.,} $|S_i|=m_i$. Here, $a$ is an appropriately chosen constant, independent of $D$ and $n$. Further, let $\lambda_i=\frac{n}{(2D)^{i-1}}$.
We will first show the for every object $o\in\dbp\cup\left\{q\right\}$, where $q$ is the query point, the following four properties of the data structure are true w.h.p. 
\begin{enumerate}
	\item \label{P1}$|S_i\cap \ball{o}{\lambda_{i+1}}|\geq 1$
	\item \label{P2}$|S_i\cap \ball{o}{\lambda_i}|\leq 4aD\log n$
	\item \label{P3}$|S_{i+1}\cap \ball{o}{\lambda_{i-1}}|\leq 16aD^3\log n$
	\item \label{P4}$|S_i\cap \ball{o}{4\lambda_i}|\geq 4aD\log n$
	\item \label{P5}$|S_{i+1}\cap \ball{o}{4\lambda_{i-1}}|\leq 64aD^3\log n$ 
\end{enumerate}
Fix an object $o$ and a level $i$. To visualize the proof, place all objects in the database on a line, such that the object $u$
with rank $r_o(u,\dbp)=r$ is located at distance $r$ from $o$ (see
figure \ref{fig:hds}). 
\begin{figure}[htbp]
	\centering
		\includegraphics[width=0.8\columnwidth,keepaspectratio]{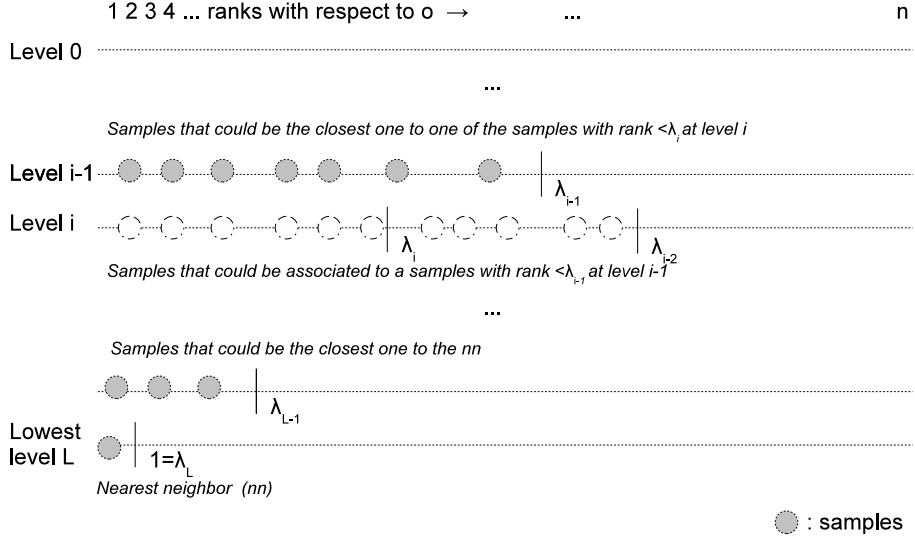}
		\caption{We place all objects on the line such that
                  the object $u$ with rank $r_o(u,\dbp)=r$ is located
                  at distance $r$ from $o$. }
		\label{fig:hds}
\end{figure}
Property \ref{P1} tells us that at least one of the samples at level $i$ will be such that its rank w.r.t. $o$ is smaller than $\lambda_{i+1}$ \emph{i.e.,} $\exists s\in S_i$ s.t. $r_o(s)\leq \lambda_{i+1}$. Clearly, by Lemma \ref{lem:tl1}, this is true with probability at least $1-\frac{1}{n^a}$ (set $m=m_i$ and $b=(2D)^i=\frac{n}{\lambda_{i+1}}$ in the lemma). Property \ref{P2} tells us that not too many objects can have rank less than $\lambda_i$ at level $i$ w.r.t. $o$. Let $c=\frac{n}{\lambda_i}=(2D)^{i-1}$. Now, by lemma \ref{lem:tl2} (set $m=m_i=a(2D)^i\log n$ and $\tau=1$), the probability that more than $2a(2D)^i\log n/(2D)^{i-1}=4aD\log n$ samples are among the $\lambda_i=\frac{n}{c}$ closest samples to $o$ is less than $2 e^{- 2aD\log n/3}=\frac{1}{n^{\Omega(a)}}$. The proof of Property \ref{P3} is identical, except that we replace $\lambda_i$ by $\lambda_{i-1}$. Then, we have $c=(2D)^{i-2}$, $\frac{2m_{i+1}}{c}=16aD^3\log n$, and the probability that $|S_{i+1}\cap \ball{o}{\lambda_{i-1}}|> 16aD^3\log n$ is smaller than $\frac{1}{n^{\Omega(a)}}$, as before. For Property \ref{P4}, we expect $8aD\log n$ objects to be sampled at level $i$ among the $4\lambda_i$ closest objects to $o$. Again, by lemma $\ref{lem:tl2}$, the probability that less than half that many objects get sampled is at most $\frac{1}{n^{\Omega(a)}}$. Finally, the proof of Property \ref{P5} is almost identical to the proof of Property \ref{P2}. By choosing $a$ large enough, we can make sure that the five properties are true for all objects and all levels w.h.p. (take the union bound over the $n$ objects and the $L=\frac{\log n}{\log 2D}$ levels). 

From now on, we assume that we are in the situation where Properties (1) to (5) are true for all objects (which is the case w.h.p.). Again, fix an object $o$. Consider a sample $s\in S_{i+1}$ such that $r_{o}(s)\leq \lambda_{i+1}$ (note that Property \ref{P1} guarantees that there is a least one such sample). Further, let $s'\in S_{i}$ be the sample at level $i$ closest to $s$ \emph{i.e.,} $s'=\min_{x\in S_{i}}r_{s}(s')$. Again, by Property \ref{P1}, we know that $r_{s}(s')\leq \lambda_{i+1}$. Hence, by the approximate triangle inequality 3 (see Section \ref{sec:def}), we have:
\mlf{
  r_{o}(s,\dbp)\leq \lambda_{i+1}\mbox{ and }r_{s}(s'),\dbp)\leq \lambda_{i+1} &\Rightarrow r_{o}(s'),\dbp)\leq 2D\lambda_{i+1}=\lambda_i
}  
Consequently, we know that the sample that is closest to $o$ at level $i+1$ will be in the bin of a sample $s'\in S_{i}$ that has rank $r_{o}(s',\dbp)\leq \lambda_i$. The algorithm associates every object $o$ to the closest sample on each level. Hence, to find that sample for object $o$ at level $i+1$, it would be sufficient to rank (w.r.t. $o$) all sample in $S_{i+1}$ that fell in the bin of a sample at level $i$ that has rank less than $\lambda_i$. Property \ref{P2} tells us that $|S_i\cap \ball{o}{\lambda_i}|\leq 4aD\log n$. Hence, by inspecting the bins of the at most $4aD\log n$ closest samples to $o$ at level $i$, and ranking the samples at level $i+1$ that fall in those bins, we are guaranteed to find the closest sample. Property \ref{P4} tells us that all of the $4aD\log n$ closest samples to $o$ at level $i$ have rank less than $8\lambda_i$. Consider a sample $s\in S_i$ such that $r_{o}(s,\dbp)\leq 8\lambda_i$ and a sample $s''\in S_{i+1}$ that falls in the bin of $s$. By property \ref{P1}, we must have $r_{s''}(s,\dbp)\leq \lambda{i+1}$. Thus, by inequality 2, we have:
\mlf{ 
r_{s''}(s,\dbp)\leq \lambda_{i+1}\mbox{ and }r_o(s,\dbp)\leq 8\lambda_{i} \Rightarrow r_o(s'',\dbp)<D(8\lambda_{i}+\lambda_{i+1})\leq 4\lambda_{i-1}
  } 
By property \ref{P5}, there are at most $O(D^3\log n)$ such samples at level $i+1$.

To summarize, at every level in the hierarchy, and for every object, we need to rank at most $O(D^3\log n)$ samples. Consequently, we need to ask at most $O(nD^3\log^2n\log\log n^{D^3})$ questions in total to rank at most $O(D^3\log(n))$ objects for every object and level. The algorithm only fails with negligible (smaller than $\frac{1}{n}$) probability if an object has no sample that falls within $\lambda_i$ at any level $i$. 
\end{proof}

\subsection{Proof of Theorem \ref{thm:ubo}}
\label{apx:ubo}
\begin{proof}
The upper bound on the number of questions to be asked in the learning phase is immediate from Theorem \ref{thm:alg1}. For each object, we need to store one identifier (the identifier of the closest object) at every level $i$ in the hierarchy, and one bit to mark it as a member of $S_i$ or not. Hence, the total memory requirements\footnote{Making the assumption that every object can be uniquely identified with $\log n$ bits} do not exceed $O(n\log^2n/\log(2D)$ bits. Finally, the properties 1-5 shown in the proof of Theorem \ref{thm:alg1} in Appendix  \ref{apx:alg1}  are also true for an external query object $q$. Hence, to find the closest object to $q$ on every level, we need to ask at most $O(D^3\log^2n\log\log n^{D^3})$ questions. In particular, the closest object at level $L=\log_{2D}(n)$ will be $q$'s nearest neighbor w.h.p. 
\end{proof}

\subsection{Proof of Lemma \ref{lem:dlb}}
\label{apx:dlb}
\begin{proof}
  Consider the configuration given in Figure \ref{fig:lbt}. We need to show that for all triples $x,y,z$, where $x,y,z \in
  \dbp\cup\left\{q\right\}$, we have $r_x(y)\leq
  D(r_z(x)+r_z(y))$. First, let us consider two nodes $x$ and $y$ such
  that $d(x,y)=d$, with $d>1$. Clearly, these two nodes must be in two
  different supernodes as the maximum distance inside a supernode is
 is strictly smaller than $1$. Further, we have $|\ball{x}{d}|\leq 4\alpha^2d$. Indeed, even if $x$ is
  in the supernode at the center of the star, there are at most $\alpha d$
  other supernodes within distance $d$. Each supernode can contain at
  most $\alpha$ nodes. Further, the query point could be within distance
  $d$ of $x$, in that case there could be at most $2d\alpha^2$ additional
  objects in the balls. On the other hand, we have $|\ball{z}{d/2}|\geq
  d\alpha/2$. Indeed, even if $z$ is placed at the end of a branch, there
  are at least $d\alpha$ supernodes within distance $d$, each containing
  $\alpha$ nodes. Hence, we have $r_x(y)\leq 4\alpha^2d \leq 2D\alpha d\leq
  D(r_z(x)+r_z(y))$ by setting $\alpha=D/2$. We have used the fact that
  $r_z(x)+r_z(y)\geq |\ball{z}{j}|+ |\ball{z}{d-j}|\geq
  2|\ball{z}{d/2}|$.
\par
If the distance is smaller than 1, then $x$ and $y$
must be inside the same supernode. In that case, we have $r_x(y)\leq \alpha
\leq 2D \leq D(r_z(x)+r_z(y))$. We can prove the other inequalities in a similar way. 
\end{proof}

\subsection{Proof of Theorem \ref{thm:lbo}}
\label{apx:lbo}
\begin{figure}[htbp]
	\centering
        \includegraphics[width=0.9\columnwidth,keepaspectratio]{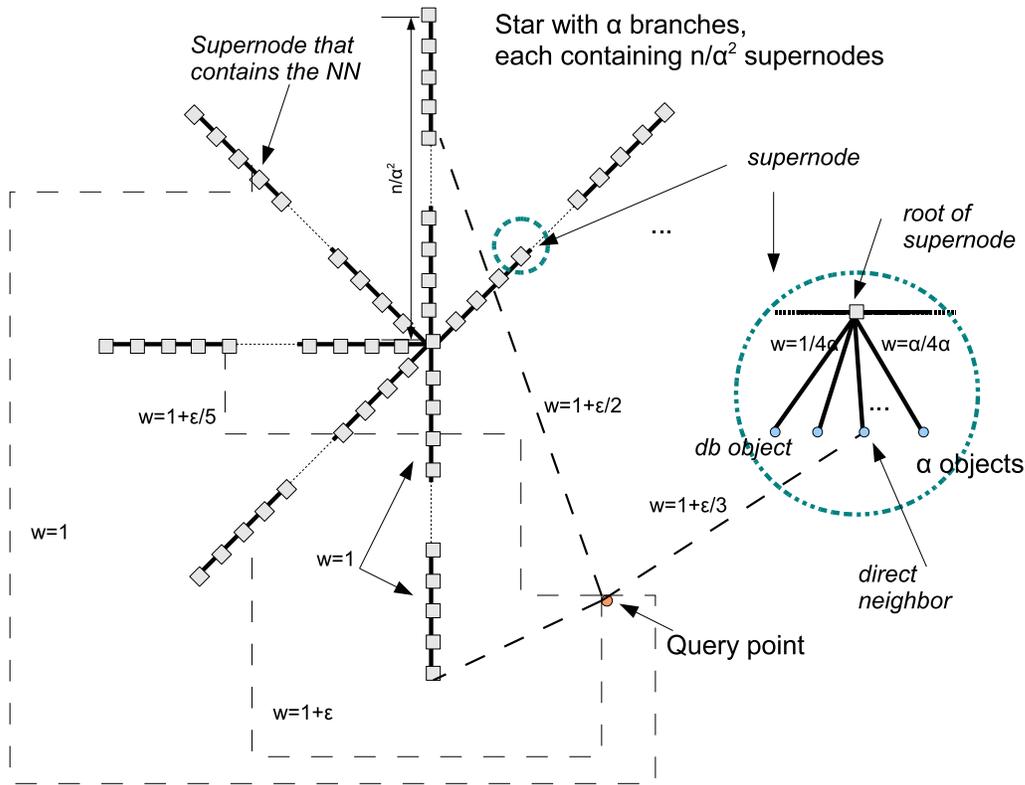}
        \caption{A graph with disorder constant $\alpha$ and shortest path distance. The graph forms a star with $\alpha$ branches. Each branch is composed of $n/\alpha^2$
          ``supernodes''. Each edge between the "roots" of the supernodes in
          the star has weight $1$ ($w$ denotes the weight in the figure).  Each supernode (see zoomed region on the right side of the figure) is in turn composed of a "root", and a smaller $(\alpha)$-ary tree (of depth 1) consisting of $\alpha$ actual database objects. The
          weights of the edges in the tree range from $1/4\alpha$ to
          $\alpha/4\alpha$. Finally, a query point is randomly connected to
          \emph{one} non-root node on \emph{each} branch of the star with edges of weights
          ranging from $1$ to $1+\epsilon$ (dashed lines), where $\epsilon\ll 1/4\alpha$. The distances on the graph are shortest path distances. There are $(\frac{n}{\alpha})^\alpha$ ways to connect the query point to the database. Further, for each such \emph{configuration}, there are $\alpha$ possible choices for the nearest neighbor (the direct neighbor which is connected to the query point with the edge of smallest weight). We assign weights to the edges connecting direct neighbors to the query point in such a way that each of the direct neighbors is equally likely to be the nearest neighbor, and each weight is different.}
	\label{fig:lbt}
\end{figure}
\begin{proof}
Consider the graph metric with shortest path distance in Figure \ref{fig:lbt}. Yao's minimax principle (see \cite{bra}) states that, for any distribution on the inputs the expected cost for the best deterministic algorithm provides a lower bound on the expected running
time of any randomized algorithm. The graph (solid lines in Figure \ref{fig:lbt}) is known. It consists of a star with $\alpha$ branches, each composed of $\frac{n}{\alpha^2}$ supernodes. Each of the supernodes in turn contains $\alpha$ database objects (\emph{i.e.,} objects in $\dbp$). Clearly, in total there are $\alpha\alpha\frac{n}{\alpha^2}=n$ objects. We know the answers to all questions of the type $\ora{a}{b}{c}$, where $a,b,c\in\dbp$. We attach a query point $q$ to that graph, and we assume that each "position" of the query point (as shown in
Fig. \ref{fig:lbt}) is equally likely. That is, the query point is attached to one (non-root) object chosen u.a.r. on each branch of the star with an edge. This object is a called a \emph{direct neighbor}. The weights of the corresponding edges are chosen between $1$ and $1+\epsilon$ in a random way as well (such that we do not have ties, and each of the direct neighbors is equally likely to be the nearest neighbor). In other words, the input distribution is uniform over all configurations. First, note that $q$'s
nearest neighbor must be one of the objects connected directly to
it \emph{i.e.,} one of the $\alpha$ direct neighbors. Indeed, let $\delta=\brak{u\in\dbp| \mbox{$u$ is a direct neighbor of $q$}}$. Then, we have $d(u,q)<d(v,q)$, when $u\in\delta$ and $v\in\dbp\backslash\delta$. Further, any of these direct neighbors could be $q$'s nearest
neighbor with equal probability. Assume that we are given for free the answers to all questions, except the questions of the type $\ora{q}{x}{y}$, where both $x,y\in\delta$. This amounts to knowing which are the direct neighbors, but not knowing anything about the ranking of those direct neighbors with respect to $q$. Indeed, by construction, all direct neighbors are closer to the query point than any other object in the database. Hence, if we used the oracle to compare a direct neighbor with another object (which is not a direct neighbor), the oracle would always answer that the direct neighbor is closer to the query point. So, we could not exclude one of the direct neighbors as the nearest neighbor (we do not learn anything about the nearest neighbor). Hence, in order to identify the
nearest neighbor, the best deterministic algorithm must at least ask $\alpha$ questions to find the
nearest neighbor among the direct neighbors (we must traverse the list of direct neighbors and ask the oracle to compare every object with the current best candidate). Consequently, we must first identify all direct neighbors, and then compare them with each other. 

Note that there are $(\frac{n}{\alpha})^\alpha$ ways to choose the direct neighbors, and that each configuration is equally likely. Identifying all direct neighbors is equivalent to knowing which of these configurations we are in. Let $X$ denote the random variable of which each outcome corresponds to a configuration. Then, the entropy of $X$ is $\log(\frac{n}{\alpha})^\alpha=\alpha\log(n/\alpha^2)+\alpha\log\alpha$ bits. The answer to every question we ask the oracle will reduce the uncertainty about which configuration we are in. In order for the probability of error $p_e$ to be equal to zero \emph{i.e.,} in order to be sure that we found the all direct neighbors, Fano's Inequality (see \cite{coverthomas}, p39) tells us that we must know at least a set of answers $A$ such that $H(X|A)=1$ bit to have the $p_e\geq 0$. 

For every branch of the star, choosing a direct neighbor u.a.r. is equivalent to choosing a supernode u.a.r., and then a direct neighbor inside that supernode u.a.r. First, assume that we know, on each branch, in which supernode the direct neighbor is located. Let us focus on one branch, and the supernode on this branch containing a direct neighbor. Denote that supernode by $\Phi$. In that case, in order to identify the direct neighbor in $\Phi$, we must ask questions of the type $\ora{q}{a}{b}$, where $a,b\in \Phi$. Asking a question where either $a$, $b$ or both are outside $\Phi$ does not tell us anything about which object is the direct neighbor, as all objects inside $\Phi$ are closer to $q$ than any object outside that supernode. Further, note that the answer to any question of the type $\ora{q}{a}{b}$, where $a,b\in\Phi$ and $d(z,b)>d(z,a)$ is $b$ only if $b$ is $q$'s direct neighbor in $\Phi$. Hence, the answer to a question of this type allows us to exclude only one object at a time\footnote{The same is true if we ask questions where $a$ and $b$ are in different supernodes. What matters is that we can only exclude one object as being a direct neighbor every time we ask a question}. Hence, for each of the $\alpha$ supernodes that contain a direct neighbor to $q$, we must ask an expected $\Omega(\alpha)$ questions to identify the direct neighbor. Knowing all the direct neighbors, when the supernodes in which they are located are known, reduces the entropy by $\alpha\log (\alpha)$ bits. Indeed, there are $\alpha$ such supernodes, and $\alpha$ choices for the direct neighbor inside each of these supernodes (\emph{i.e.,} if we fix the supernodes containing the direct neighbors, there are $\alpha^\alpha$ ways to choose the direct neighbors). As every question only excludes one object inside a supernode as direct neighbor, in total we must ask $\Omega(\alpha^2)$ questions to the oracle. 

Let us now remove the assumption that we know which supernodes contain a direct neighbor. There are $(\frac{n}{\alpha^2})^\alpha$ ways to choose the supernodes that contain the direct neighbors. The entropy for this random choice is consequently $\alpha\log(n/\alpha^2)$ bits. Thus, at best, we need to ask $\alpha\log(n/\alpha^2)$ questions (in the best case each question reduces the number of possible configurations by 2) in order to know in which supernodes the direct neighbors are located. In total, we consequently need to ask at least an expected $\Omega(\alpha\log \frac{n}{\alpha^2}+\alpha^2)$ questions, to reduce the entropy by $\log(\frac{n}{\alpha})^\alpha=\alpha\log(n/\alpha^2)+\alpha\log\alpha$ bits and having $p_e\geq 0$. By letting $\alpha=\Theta(D)$, we obtain the claim.   
\end{proof}

\subsection{Proof of Lemma \ref{lem:als}}
\label{apx:als}
\begin{proof}
The result follows directly from the approximate triangle inequality (see Definition \ref{apxti}). The lower bound follows from inequality 3 for (1) and inequality 2 for (2). The upper bound follows from inequality 2 for (1) and inequality 3 for (2).
\end{proof}

\subsection{Proof of Theorem \ref{thm:ann}}
\label{apx:ann}
\begin{proof}
	During a learning phase we sample $m$ objects
  $S=\left\{s_1,...s_m\right\}$ u.a.r in $\dbp$, and rank all other
  objects with respect to the objects in $S$\emph{i.e.,} $\forall s\in S, u\in\dbp$, we compute $r_s(u,\dbp)$ by querying the oracle (this can be done by asking $O(mn\log n)$ questions). 
  In the search phase, we start by finding the point in $S$ closest to
  $q$, that is we want to find $x=argmin_{s\in S}r_q(s)$. This can be
  done in $m$ steps by traversing the list of objects in $S$
  sequentially and storing the closest element seen so far. In
  particular, for every object in $S$, we ask the oracle whether it is
  closer to $q$ than the current minimum, and if so it becomes the new
  minimum. Then, using binary search, we can find $j'=r_x(q)$
  (\emph{i.e.,} we ask the oracle whether $q$ is closer to $x$ than
  the element $y$ such that $r_x(y)=n/2$, and then apply this process
  recursively on the new "interval"). Now, given that in the
  learning phase we sample $m$ objects u.a.r. in $\dbp$, we know
  that $\pr{r_q(x)=j}=\frac{m}{n}(1-\frac{j}{n})^m$. Further, we know
  by triangle inequality that $\frac{j}{D}\leq j' \leq Dj$. Hence, by
  Lemma \ref{lem:als}, all objects $o$ such that $r_q(o,\dbp)<R$ must
  lie in an annulus centered at $x$ such that $\frac{j}{D^2}-R\leq
  r_x(o)\leq D^2 j+DR$ (see Figure \ref{fig:ans}). 
  \begin{figure}[htbp]
	\centering
		\includegraphics[width=0.6\columnwidth,keepaspectratio]{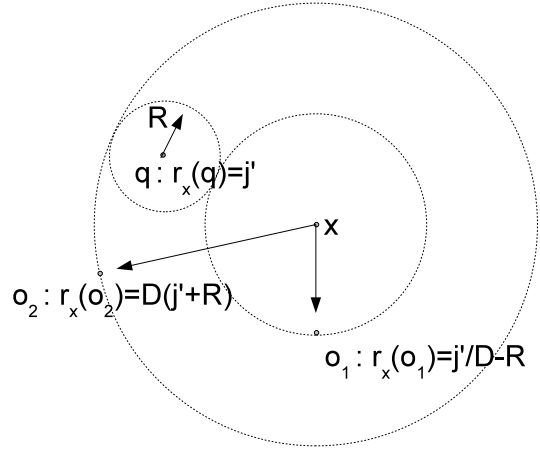}
		\caption{The $R$ nearest neighbors of $q$ must lie in an annulus around $x$.}
	\label{fig:ans}
\end{figure}
  This annulus contains at most
  $(D+1)R+j(D^2-\frac{1}{D^2})<(D+1)R+j(D^2)$ objects, of which $R$
  are the $R$ nearest neighbors of $q$. Hence, by sampling
  $(D+1)+\frac{j(D^2)}{R}$ times, we will retrieve an $R$-nearest
  neighbor with constant probability. Thus, the expected number of
  times we need to sample is
  $\sum_{j=1}^{n}\frac{m}{n}(1-\frac{j}{n})^m
  ((D+1)+\frac{j(D^2)}{R})\leq (D+1)+\frac{nD^2}{mR}$. For every
  sample, we ask the oracle if it is closer to $q$ than the currently
  closest sampled point. If so, we store this point, else we delete
  it.
\end{proof}

\subsection{Proof of Lemma \ref{lem:dia}}
\label{apx:dia}
\begin{proof}
Clearly, if $|S|=n$, for any pair of point $u$ and $v$, we have $r_u(v,S)\leq n$. Hence, $r_u(v)+r_v(u)\leq 2n$ for all $u,v$. In case the distances are symmetric in the hidden space, we can rank the distances from the smallest distance to the largest distance. Consider the pair $v,w$, such that $d(v,w)=d(w,v)>d(i,j)$, for all $i,j$. Then, we clearly have $r_v(w)=n$ and $r_w(v)=n$, since there cannot be any point further away from $v$ than $w$, and vice-versa.   
\end{proof}

\subsection{Proof of Lemma \ref{thm:btp}}
\label{apx:btp}
\begin{proof}
We need to compute the probability that $k$ is such that $\frac{4D}{n}k^2+2n-2k< \epsilon 2n$, or equivalently $\frac{2D}{n}k^2-k+(1-\epsilon)n<0$. Solving for $k$, we obtain $k=\frac{1\pm \sqrt{1-8D(1-\epsilon)}}{\frac{4D}{n}}=\frac{n}{4D}\pm\frac{n}{4D}\sqrt{1-8D(1-\epsilon)}$. Hence, the number of values of $k$ for which the above condition is fulfilled is  $|\frac{n}{4D}+\frac{n}{4D}\sqrt{1-8D(1-\epsilon)}-\frac{n}{4D}+\frac{n}{4D}\sqrt{1-8D(1-\epsilon)}|=\frac{n}{2D}\sqrt{1-8D(1-\epsilon)}$. As we choose $k$ u.a.r. from $n$ values, we have $\pr{\tilde{\Delta}_S\leq  \epsilon 2n}=\frac{1}{2D}\sqrt{1-8D(1-\epsilon)}$
\end{proof}

\subsection{Depth of binary tree}
\label{apx:dbt}
\begin{proof}
Let $\delta<0.5$ be a constant independent of $n,D$. Consider a particular path in the binary tree from the root to a leaf. Let $n_i$ denote the number if objects in the set at level $i$ and $k_i$ the rank of $x_2$ w.r.t $x_1$ (\emph{i.e.,} $r_{x_1}(x_2)$) chosen at level $i$. Let $X_i=1$ if $\delta n_i\leq k_i \leq (1-\delta) n_i$. As $k_i$ is distributed u.a.r. in $1,..,n_i$, we have $\pr{X_i=1}=1-2\delta$. If $X_i=1$, the number of objects is reduced by a factor at least $1-\delta$ at this level \emph{i.e.,} $\max\left\{|S_0|,|S_1|\right\}<(1-\delta)n_i$. As there are $n$ objects in total, we can not reduce the number of objects by a factor $(1-\delta)$ more than $s=\frac{\log(n)}{\log(1/(1-\delta))}$ times. In $m$ levels on the path, the expected number of times we expect $X_i$ to be equal to $1$ is $\mu=(1-2\delta)m$. If we set $m=\frac{2as}{(1-2\delta)}$, for some constant $a>1$ we have 
$$
\pr{\sum_{j=1}^{m}X_j<\mu/2=as}<O(1/poly(n))
$$
By the Chernoff bound. There are at most $n$ paths (ones per leaf). Taking the union bound over these paths, we obtain the claim.
\end{proof}

\subsection{Proof of theorem \ref{thm:rhf}}
\label{apx:rhf}
\begin{proof}
First, we compute the probability that the hash function $h$ is different for two objects $u$ and $q$.
\mlf{
p&=\pr{h(u)\neq h(q)}\\
&=\sum_{i,j\in\dbp}\pr{h(u)\neq h(q)|x_1=i,x_2=j}\pr{x_1=i,x_2=j}\\
&=\frac{1}{n}\sum_{i} \pr{r_{x_1}(x_2)\in \left[r_{x_1}(q),r_{x_1}(u)\right]|x_1=i }\\
&=\frac{1}{n^2}\sum_{i}|r_i(u)-r_i(q)|\\
&=\frac{1}{n^2}||\rho_{u}-\rho_{q}||_1
}
Hence, we have $$\pr{h(u)= h(q)|r_q(u)\leq r}=1-\frac{1}{n^2}||\rho_{u}-\rho_{q}||_1\geq 1-\frac{f(r)}{n^2}$$, and similarly $$\pr{h(v)= h(q)|r_q(v)\geq (1+\epsilon)}=1-\frac{1}{n^2}||\rho_{v}-\rho_{q}||_1\leq 1-\frac{f((1+\epsilon)r)}{n^2\gamma})$$
\end{proof}

\subsection{Proof of Corollary \ref{cor:rsh}}
\label{apx:rsh}
\begin{proof}
The proof is analogous to the proof for locality-sensitive hashing for binary vectors provided in \cite{lshash}. More precisely, for an $(r,R,p,P)$-rank sensitive hashing scheme, retrieving one of the $R$ nearest neighbor of a query point $q$ will requires $O(n^\theta)$ evaluations of the hash function. $\theta$ is defined as $\frac{\log\frac{1}{p}}{\log\frac{p}{P}}$. It can be shown that $\theta\leq \frac{1}{\frac{1+\epsilon}{\gamma}-1}=O(\frac{\gamma}{1+\epsilon})$. Indeed, the probabilities $p$ and $P$ take the same values as if we hashed binary vectors of dimension $n^2c$, and let $r'=r$, and $(1+\epsilon')r'=(1+\epsilon)r/\gamma$. Then, $\theta\leq \frac{1}{\epsilon'}$  
\end{proof}

\newpage
\subsection{Numerical example for rank distortion}
\label{apx:nex}
\begin{figure}[htbp]
\centering
\includegraphics[width=1\columnwidth,keepaspectratio]{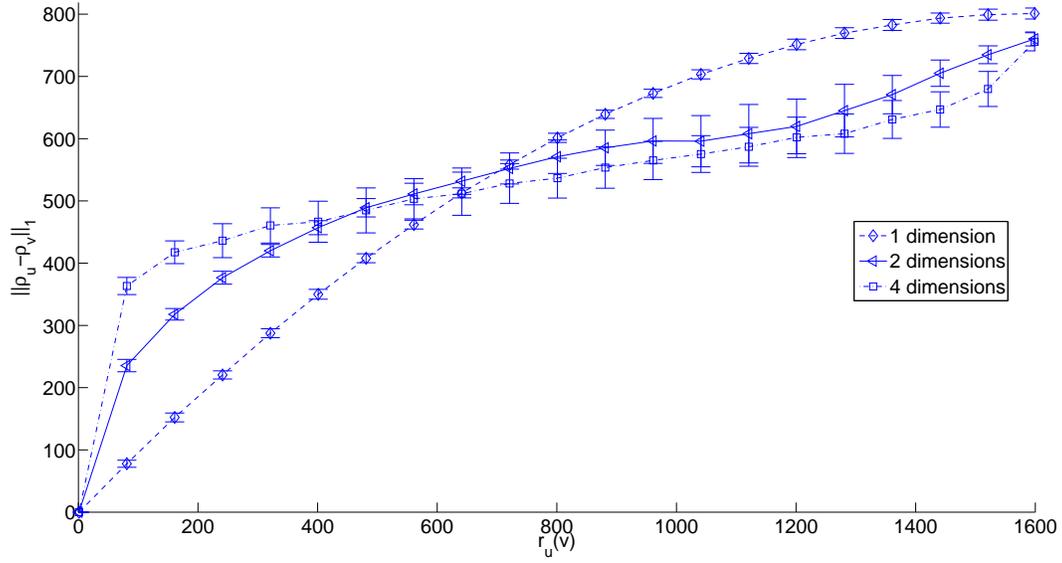}
\caption{The hidden space consists of 1600 points distributed u.a.r on $\left[0,1\right]^d$, where $d=1,2,4$. To avoid border effects, we compute distances with wrap-around. We plot the $||\rho_u-\rho_v||$ against $r_u(v)$, for a fixed $u$. The results are averaged over $100$ samples and the error bars correspond to the standard deviation. Note that the slope is first steep and then linear. Such a function is appropriate for RSH, as the function $f$ increases monotonically. Further, the fact that we have a steep slope for small values of $R$ make those spaces particularly attractive. Indeed, this implies that $P$ decreases rapidly (so we can search for $R$-nearest neighbors, even for small $R$), and $p$ is sufficiently large for small values of $r$. This example shows that for homogeneous spaces, the rank distortion function is such that we can perform RSH efficiently.}
\label{fig:rdu}
\end{figure}

\end{document}